\documentclass[
prd
,showpacs ,amssymb,nobibnotes,aps,eqsecnum]{revtex4}
\usepackage{graphicx}
\input{epsf}

\usepackage{amsmath,amssymb}
\usepackage{bm}

\newcommand{\dalm}{\kern1pt\vbox{\hrule height 0.9pt\hbox{\vrule width 0.9pt
\hskip 2.5pt\vbox{\vskip 5.5pt}\hskip 3pt\vrule width 0.3pt}\hrule height 0.3pt}
\kern1pt}

\newcommand{\gsim}{\, \raisebox{-0.8ex}{$\stackrel{\textstyle >}{\sim}$ }}

\begin{document}



\title{Slowly Rotating Relativistic Stars in Tensor-Vector-Scalar Theory}

\author{Hajime Sotani$^{1,2}$} \email{sotani@astro.auth.gr}
\affiliation{
$^1$ Theoretical Astrophysics, University of T\"{u}bingen,
Auf der Morgenstelle 10, T\"{u}bingen 72076, Germany \\
$^2$ Advanced Research Institute for Science and Engineering, Waseda University,
Shinjukuku, Tokyo 169-8555, Japan
}

\date{\today}

\begin{abstract}
In order to examine the rotational effect around neutron star in tensor-vector-scalar (TeVeS) theory, we consider the slowly rotating relativistic stars with a uniform angular velocity. As a result, we find that similar to the case in general relativity (GR), the angular momentum is proportional to the angular velocity. Additionally, as the value of coupling constant $K$ becomes higher, the frame dragging in TeVeS becomes quite different distribution from that in GR, where we can also see the deviation even in the interior of star. While with smaller value of $K$, although the frame dragging approaches to that expected in GR, the induced vector field due to the rotation does not vanish and still exists. Thus, through the observations associated with relativistic object, one could be possible to distinguish the gravitational theory in strong field regime even in the case that the value of coupling constant $K$ is quite small.
\end{abstract}

%
\maketitle
\section{Introduction}
\label{sec:I}

In the weak gravitational field such as solar system, there are many experiments and the validity of general relativity (GR) has been shown, while the gravitational theory in the strong field regime are still unconstrained by the observations. However, the development of technology will be possible to observe compact objects with high accuracy and those observations could be used as a direct test of the gravitational theory in strong field regime \cite{Psaltis2008}. In fact, there are attempts to test the gravitational theory by using surface atomic line redshifts \cite{DeDeo2003} or gravitational waves emitted from the neutron stars \cite{Sotani2004}. In these works, they suggest the possibility to distinguish the scalar-tensor theory proposed in \cite{Damour1992} from GR.

As an alternative gravitational theory, the tensor-vector-scalar (TeVeS) theory has attracted considerable attention, which is proposed originally by Bekenstein \cite{Bekenstein2004} as a covariant theory for modified Newtonian dynamics \cite{Milgrom1983} (see \cite{Skordis2009} for review of TeVeS). The advantage to adopt this theory is possible to explain the galaxy rotational curves and the Tully-Fisher law without the existence of dark matter \cite{Bekenstein2004}. Additionally, TeVeS has successfully explained not only strong gravitational lensing \cite{Chen2006} but also the galaxy distribution through an evolving Universe without cold dark matter \cite{Dodelson2006}. In the strong gravitational region of TeVeS, the Schwarzschild solution was found by Giannios \cite{Giannios2005}, and the Reissner-Nordstr\"{o}m solution was found by Sagi and Bekenstein \cite{Sagi2008}. Furthermore, Lasky et al. derived the Tolman-Oppenheimer-Volkoff (TOV) equations in TeVeS and they produced the static, spherically symmetric neutron star models in TeVeS \cite{Paul2008}.

Recently, there are some suggestions to distinguish TeVeS from GR by using some observations, i.e., with the redshift of the atomic spectral lines radiated from the surface of neutron star \cite{Paul2008},  with the Shapiro delays of gravitational waves and photons or neutrinos \cite{Desai2008}, and with the emitted gravitational waves from neutron stars \cite{Sotani2009a}. In this article, we examine the different way to distinguish TeVeS from GR, i.e., the rotational effect around neutron stars. For this purpose, we consider the slowly rotating neutron stars with a uniform angular velocity. The analysis of slowly rotating neutron star in GR has originally done by Hartle \cite{Hartle1967}, and subsequently many works have done in this field. 
Actually, taking into account the rotational effect is more natural and more important. For example, due to the rotational effect, the new oscillation family, i.e., so-called $r$ mode, could be excited \cite{Miltos2008,Erich2009}.  So, this article would become a first step to consider the rotational effect in more complicated system in TeVeS.

This article is organized as follows. In the next section, we describe the fundamental parts of TeVeS and the equations to produce the non-rotating relativistic stellar models in TeVeS,  which are corresponding to TOV equations in GR. In section \ref{sec:III}, we derive the equations to describe the slowly rotating relativistic stars. In order to discuss about the slowly rotating star models in TeVeS, as mentioned later, two variables are required, i.e., one is the variable shown the frame dragging and the other is the vector field induced by the rotation. In this section,  we also show the numerical results for many stellar models. At last, we make a conclusion in section \ref{sec:IV}.
In this article, we adopt the unit of $c=G=1$, where $c$
and $G$ denote the speed of light and the gravitational constant, respectively, and
the metric signature is $(-,+,+,+)$.

\section{Stellar Models in TeVeS}
\label{sec:II}
\subsection{TeVeS}
\label{sec:II-1}

In this section, we briefly describe the fundamental parts of TeVeS, which are necessary for the calculations in this article (see for the details of this theory in \cite{Bekenstein2004}).
TeVeS is constructed with three dynamical gravitational fields, i.e., an Einstein metric $g_{\mu\nu}$,  a timelike 4-vector field ${\cal U}^\mu$, and a scalar field $\varphi$, in addition to a nondynamical scalar field $\sigma$. The vector field fulfills the normalization condition, $g_{\mu\nu}{\cal U}^\mu{\cal U}^\nu=-1$, and the physical metric is given by
\begin{gather}
 \tilde{g}_{\mu\nu} = e^{-2\varphi}g_{\mu\nu} - 2{\cal U}_\mu{\cal U}_\nu\sinh(2\varphi),
\end{gather}
along which usual matter such as electromagnetic fields propagate. Hereafter, all quantities in the physical frame are denoted with a tilde, and any quantity without a tilde is in the Einstein frame.  
The total action of TeVeS, $S$, contains contributions from the three dynamical fields and a matter
contribution, i.e., the Einstein-Hilbert action, $S_g$, the vector field's action, $S_v$, the scalar's action, $S_s$, and the matter action, $S_m$. These four parts of action can be written down as
\begin{align}
 S_g =& \frac{1}{16 \pi G} \int{g^{\alpha\beta}R_{\alpha\beta}\sqrt{-g}d^4x}, \\
 S_v =& -\frac{K}{32\pi G}\int{\left[\left(g^{\alpha\beta}g^{\mu\nu}{\cal U}_{[\alpha,\mu]}{\cal U}_{[\beta,\nu]}\right)-\frac{2\lambda}{K}\left(g^{\mu\nu}{\cal U}_\mu{\cal U}_\nu+1\right)\right]\sqrt{-g}d^4x}, \\
 S_s =& -\frac{1}{2k^2\ell^2G}\int{F(k\ell^2h^{\alpha\beta}\varphi_{,\alpha}\varphi_{,\beta})\sqrt{-g}d^4x}, \\
 S_m =& \int{{\cal L}\left(\tilde{g}_{\mu\nu},f^\alpha,f^\alpha_{|\mu},\cdots\right)\sqrt{-\tilde{g}}d^4x},
\end{align}
where ${\cal U}_{[\alpha,\beta]}\equiv{\cal U}_{\alpha,\beta}-{\cal U}_{\beta,\alpha}$,  $h^{\alpha\beta}\equiv g^{\alpha\beta}-{\cal U}^\alpha{\cal U}^\beta$, $k$ and $K$ are positive dimensionless parameters associated with the scalar and vector fields respectively, $F$ is a dimensionless free function, $\ell$ is a constant length scale, $\lambda$ is a spacetime dependent Lagrange multiplier, $f$ denotes the field variables collectively, and the covariant derivative denoted by $|$ is taken with respect to $\tilde{g}_{\alpha\beta}$ (see \cite{Bekenstein2004} for details).

Although the free function, $F$, is not predicted by the theory, as mentioned later our results in this article are independent of this function and therefore our results are also independent from the value of $\ell$. The scalar field coupling, $k$, has been constrained by Bekenstein \cite{Bekenstein2004} to be $k\simeq 0.03$ by using planetary motions in the outer solar system. While, the restrictions on the vector field coupling, $K$, are less severe, but probably one should consider the range of $0<K<2$, because for $K>2$ one can show that the stellar pressure diverges from the stellar center outward and one can not construct the stellar models \cite{Paul2008}. Additionally, the Lagrange multiplier, $\lambda$, can be described as a function of field variables from the vector equation with the normalization condition (see Eq.(\ref{lambda00})). Furthermore, we should notice about the propagation speed. That is, Bekenstein showed that TeVeS allows for superluminal propagation of tensor, vector, and scalar perturbations when $\varphi<0$ \cite{Bekenstein2004}. Using perturbations of the various fields in the physical frame, this statement was shown and was independent of the matter content of the model. Thus, in order to avoid the violating causality, the scalar field should not be negative. Namely, one should construct the neutron star model keeping that the scalar field be everywhere greater than or equal to zero. With this condition, the cosmological value of scalar field, $\varphi_c$, can be constrained as $\varphi_c\gsim 0.001$ \cite{Paul2008}.

By varying the total action, 
with respect to $g^{\mu\nu}$, one can obtain the field equations for the tensor field
\begin{equation}
 G_{\mu\nu} = 8\pi G \left[\tilde{T}_{\mu\nu}+\left(1-e^{-4\varphi}\right){\cal U}^\alpha
               \tilde{T}_{\alpha(\mu}{\cal U}_{\nu)}+\tau_{\mu\nu}\right]+\Theta_{\mu\nu},
               \label{Einstein}
\end{equation}
where $\tilde{T}_{\mu\nu}$ is the energy-momentum tensor in the physical frame, $\tilde{T}_{\alpha(\mu}{\cal U}_{\nu)}\equiv\tilde{T}_{\alpha\mu}{\cal U}_{\nu}
+\tilde{T}_{\alpha\nu}{\cal U}_{\mu}$ and $G_{\mu\nu}$ is the Einstein tensor
in the Einstein frame.  Conservation of energy-momentum is therefore given in
the physical frame as $\tilde{\nabla}_\mu \tilde{T}^{\mu\nu}=0$.
The other sources in Eq.(\ref{Einstein}) are given by
\begin{align}
 \tau_{\mu\nu} =& \sigma^2 \bigg[\varphi_{,\mu}\varphi_{,\nu}-\frac{1}{2}g^{\alpha\beta}
     \varphi_{,\alpha}\varphi_{,\beta}g_{\mu\nu} - \frac{G\sigma^2}{4\ell^2}F(kG\sigma^2)
     g_{\mu\nu}
     - {\cal U}^\alpha \varphi_{,\alpha}\left({\cal U}_{(\mu}\varphi_{,\nu)}
     -\frac{1}{2}{\cal U}^\beta\varphi_{,\beta}g_{\mu\nu}\right)\bigg], \label{tau} \\
 \Theta_{\mu\nu} =& K\left(g^{\alpha\beta}{\cal U}_{[\alpha,\mu]}{\cal U}_{[\beta,\nu]}
     - \frac{1}{4}g^{\gamma\delta}g^{\alpha\beta}{\cal U}_{[\gamma,\alpha]}{\cal U}_{[\delta,\beta]}
     g_{\mu\nu}\right)
     - \lambda {\cal U}_{\mu}{\cal U}_{\nu}.
\end{align}
In the same fashion, by varying $S$ with respect to ${\cal U}_\mu$ and $\varphi$,
one obtains the field equations for the vector and scalar fields;
\begin{gather}
 K{{\cal U}^{[\alpha;\beta]}}_{;\beta} + \lambda {\cal U}^\alpha + 8\pi G\sigma^2{\cal U}^\beta
      \varphi_{,\beta}g^{\alpha\gamma}\varphi_{,\gamma}
      = 8\pi G \left(1-e^{-4\varphi}\right)g^{\alpha\mu}
      {\cal U}^\beta\tilde{T}_{\mu\beta}, \label{vector} \\
 \left[\mu(k\ell^2h^{\mu\nu}\varphi_{,\mu}\varphi_{,\nu})h^{\alpha\beta}\varphi_{,\alpha}\right]_{;\beta}
      = kG\left[g^{\alpha\beta}+ \left(1+e^{-4\varphi}\right){\cal U}^\alpha{\cal U}^\beta\right]
      \tilde{T}_{\alpha\beta}, \label{scalar}
\end{gather}
where $\mu(x)$ is a function defined by $2\mu F(\mu) + \mu^2dF(\mu)/d\mu = -2x$.
With this function $\mu$, the nondynamical scalar field $\sigma$ is determined by
\begin{equation}
 kG\sigma^2 = \mu(k\ell^2h^{\alpha\beta}\varphi_{,\alpha}\varphi_{,\beta}). \label{scalar1}
\end{equation}
At last, the field equations in TeVeS are Eqs. (\ref{Einstein}) and (\ref{vector}) -- (\ref{scalar1}).

Now with the normalization condition for ${\cal U}^\mu$,
from the vector equation (\ref{vector}) one can calculate the Lagrange multiplier $\lambda$;
\begin{equation}
 \lambda = K {\cal U}_\alpha{{\cal U}^{[\alpha;\beta]}}_{;\beta} + \frac{8\pi}{k}
           {\cal U}^{\alpha}{\cal U}^{\beta}\varphi_{,\alpha}\varphi_{,\beta}
           - 8\pi G\left(1-e^{-4\varphi}\right){\cal U}^\alpha{\cal U}^\beta\tilde{T}_{\alpha\beta}.
           \label{lambda00}
\end{equation}
Since it has been shown in the strong-field limit that $\mu=1$ is an excellent approximation \cite{Giannios2005,Sagi2008}, 
we concentrate on the case with $\mu=1$ \footnote{
When one will consider the phenomena on cosmological scales, the assumption that $\mu=1$ might not be a good choice \cite{Bekenstein2004}.}.
This implies from Eq. (\ref{scalar1}) that $\sigma^2 = 1/(kG)$.
Moreover, 
since one can show that with $\mu=1$ the contribution of $F$ to the field equations
vanishes \cite{Bekenstein2004, Giannios2005, Sagi2008}, our
results are independent of this function and we drop it from the remaining discussion.
At the end of this section, we should mention about the assumption that $\mu=1$. As described later, to determine the frame dragging effect we should impose the asymptotical flatness. Strictly speaking, in the asymptotic region, the assumption that $\mu=1$ might not be good, but we set the numerical boundary to be $r=300M$ in this article and the results are independent from the position of numerical boundary if the numerical boundary set to be far from $r=300M$. We consider that this numerical boundary might not be so far from star and as a first step we assume that $\mu=1$ in the whole numerical region.

\subsection{Non-rotating Relativistic Stellar Models in TeVeS}
\label{sec:II-2}

The equilibrium configurations of non-rotating relativistic stars in TeVeS have been investigated by Lasky et al. \cite{Paul2008}, where they derived the Tolman-Oppenheimer-Volkoff (TOV) equations in TeVeS.  In this subsection, we briefly show the TOV equations.
Static, spherically symmetric metric can be expressed as
\begin{equation}
 ds^2 = g_{\alpha\beta} dx^\alpha dx^\beta
      = -e^{\nu(r)} dt^2 + e^{\zeta(r)} dr^2 + r^2 d\theta^2 + r^2 \sin^2\theta d\phi^2
\end{equation}
where $e^{-\zeta} = 1-2m(r)/r$. In general, the vector field for a static, spherically symmetric spacetime can be described as ${\cal U}^\mu=\left({\cal U}^t(r),{\cal U}^r(r),0,0\right)$. But
Giannios \cite{Giannios2005} showed that in vacuum, the parameterized post-Newtonian (PPN) coefficients for a spherically symmetric, static spacetime with a non-zero ${\cal U}^{r}$ can violate observational restrictions. Thus in this article we only consider the case where ${\cal U}^r=0$, which is the same assumption in \cite{Paul2008,Sotani2009a}.  In
this case, the vector field can be fully determined from the normalization
condition, such as ${\cal U}^\mu = \left(e^{-\nu/2},0,0,0\right)$.  
With this vector field, the physical metric is
\begin{equation}
 d\tilde{s}^2 = \tilde{g}_{\alpha\beta}dx^\alpha dx^\beta
              = -e^{\nu+2\varphi}dt^2 + e^{\zeta-2\varphi}dr^2
              + e^{-2\varphi}r^2 \left(d\theta^2 + \sin^2\theta d\phi^2\right) \label{phys-metric}
\end{equation}
We further assume the stellar matter content to be a perfect fluid, i.e.,
$\tilde{T}_{\mu\nu} = (\tilde{\rho} + \tilde{P})\tilde{u}_\mu\tilde{u}_\nu
                    + \tilde{P}\tilde{g}_{\mu\nu}$,
where $\tilde{\rho}$ is the energy density, $\tilde{P}$ is the pressure, and $\tilde{u}_\mu$ is a fluid four-velocity given by $\tilde{u}_\mu = e^\varphi{\cal U}_\mu$.
Then one can show that the full system of equations reduces to
\begin{gather}
 \left(1-\frac{K}{2}\right)m'
            = \frac{Km}{2r} + 4\pi G r^2 e^{-2\varphi}\left(\tilde{\rho} + 2K\tilde{P}\right)
            + \left[\frac{2\pi r^2}{k}\psi^2 - \frac{Kr\nu'}{4}\left(1+\frac{r\nu'}{4}\right)\right]
              e^{-\zeta}, \\
 \frac{Kr}{4}\nu' = -1 + \left[1+K\left(\frac{4\pi Gr^3\tilde{P}e^{-2\varphi}+m}{r-2m}
            + \frac{2\pi r^2}{k}\psi^2\right)\right]^{1/2}, \\
 \tilde{P}' = -\frac{\tilde{P} + \tilde{\rho}}{2}(2\psi + \nu'), \label{eq:tov} \\
 \varphi'   = \psi, \\
 \psi'      = \left[\frac{m' r - m}{r(r-2m)}-\frac{\nu'}{2} - \frac{2}{r}\right]\psi
            + kGe^{-2\varphi+\zeta}\left(\tilde{\rho} + 3\tilde{P}\right),
\end{gather}
where a prime denotes a derivative with respect to $r$.
Adding an equation of state (EOS), this system of equations can be closed.
The stellar radius in physical frame, $R$, is determined by $R\equiv e^{-\varphi(r_s)}r_s$,
where $r_s$ is the position of the stellar surface defined as the point where $\tilde{P}=0$.
As additional physical properties, one can introduce the total Arnowitt-Deser-Misner (ADM) mass, $M_{\rm ADM}$, and the scalar mass, $M_\varphi$, which are defined by
\begin{gather}
      M_{\rm ADM} =\left(m_{\infty}+\frac{kGM_{\varphi}}{4\pi}\right)e^{-\varphi_{c}}, \\
      M_\varphi  = 4\pi\int_0^{r}r^2\left(\tilde{\rho}+3\tilde{P}\right)e^{(\nu+\zeta)/2-2\varphi}dr,
\end{gather}
where $m_{\infty}$ and $\varphi_c$ are the mass function evaluated at radial infinity and the cosmological value of the scalar field, respectively.

In particular, in this article, we adopt the same EOS as in \cite{Sotani2004,Paul2008,Sotani2009a}, which are polytropic ones derived by fitting functions to tabulated data of realistic EOS known as  EOS A (soft EOS) and EOS II (intermediate EOS). In practice, the adopted EOS can be described as
\begin{gather}
 \tilde{P} = {\cal K}n_0m_b\left(\frac{\tilde{n}}{m_b}\right)^\Gamma, \\
 \tilde{\rho} = \tilde{n}m_b + \frac{\tilde{P}}{\Gamma-1}, \\
 m_b = 1.66\times 10^{-24} \ {\rm g}, \\
 n_0 = 0.1 \ {\rm fm}^{-3},
\end{gather}
where $\Gamma=2.46$ and ${\cal K}=0.00936$ for EOS A and $\Gamma=2.34$ and ${\cal K}=0.0195$ for EOS II. Additionally, we adopt the values of parameters $k$ and $\varphi_c$ as $k=0.03$ and $\varphi_c=0.003$, which are same choices as in \cite{Sotani2009a}, while the value of $K$ is considered in the range of $0<K<2$ (see \cite{Paul2008} for discussion about the range of value of $K$).

\section{Slowly Rotating Relativistic Stars in TeVeS}
\label{sec:III}
\subsection{Rotational Dragging}
\label{sec:III-1}

In this article, we consider a slowly rotating relativistic stellar models with a uniform angular velocity $\tilde{\Omega}$, where we assume to keep only the linear effects in the angular velocity (see in \cite{Hartle1967} for the discussion about slowly rotating relativistic stars in GR). Here we put the rotational axis to be $\theta=0$. In this case, the star is still spherical because the deformation due to the rotation is of the order $\tilde{\Omega}^2$, and
the metric in physical frame is given by
\begin{equation}
 d\tilde{s}^2 =  -e^{\nu+2\varphi} dt^2 + e^{\zeta-2\varphi} dr^2 + r^2 e^{-2\varphi} \left(d\theta^2 + \sin^2\theta d\phi^2\right)
      - 2\omega r^2 e^{-2\varphi}\sin^2\theta dt d\phi,
\end{equation}
where the last term in the right hand side corresponds to the rotational effect and
in general $\omega\sim{\cal O}(\tilde{\Omega})$ can be expressed as
\begin{equation}
 \omega(r,\theta) = -\frac{\omega(r)}{\sin\theta}\partial_\theta P_\ell,
\end{equation}
where $P_\ell=P_\ell(\cos\theta)$ is the Legendre polynomial of order $\ell$.
While, the rotational effects in the other components of metric become of the order $\tilde{\Omega}^2$,
because those should behave in the same way under a reversal in the direction of rotation as under a reversal in the direction of time. Similarly, the deviations of pressure, density, and scalar field due to the rotation should be of the order $\tilde{\Omega}^2$. Hereafter, in order to distinguish the effects due to the rotation, the deviation from the background properties would be expressed by using variables with $\delta$.  Then the fluid velocity in physical frame of the order
$\tilde{\Omega}$ can be described as
\begin{equation}
 \delta \tilde{u}^\mu = (0,0,0,\tilde{\Omega}\tilde{u}^t),
\end{equation}
where $\tilde{u}^t$ is $t$-component of four velocity in the case without rotation, i.e., $\tilde{u}^t=e^{-\varphi-\nu/2}$, and non-zero component of the energy-momentum tensor of the order $\tilde{\Omega}$ is only $\delta\tilde{T_{t\phi}}$, which are given by
\begin{align}
 \delta\tilde{T}_{t\phi} = -\left(\tilde{\rho}+\tilde{P}\right)r^2e^{-2\varphi}\left(\omega\sin\theta\partial_\theta P_\ell+\tilde{\Omega}\sin^2\theta\right) + \tilde{P}\omega r^2e^{-2\varphi}\sin\theta\partial_\theta P_\ell.
\end{align}
On the other hand, in Einstein frame the deviation of metric from the spherical symmetry is determined by
\begin{equation}
 \delta g_{\mu\nu} = e^{2\varphi}\delta \tilde{g}_{\mu\nu} + \left(e^{4\varphi}-1\right)\left(\delta {\cal U}_\mu{\cal U}_\nu + {\cal U}_\mu\delta{\cal U}_\nu\right).
\end{equation}
By using the normalization condition for the vector field, one can show that $\delta {\cal U}^t=\delta {\cal U}_t=0$. So the deviation of vector field from the spherical symmetry can be described as
 \begin{equation}
  \delta {\cal U}^\mu=\left(0, W(r)P_\ell, V(r)\partial_\theta P_\ell, \frac{{\cal V}(r)}{\sin\theta}\partial_\theta P_\ell\right).
 \end{equation}
However, the variables of $W$ and $V$ should be of the order $\tilde{\Omega}^2$ because these components are corresponding to the polar parity.
Then  the non-zero component of metric in Einstein frame of the order $\tilde{\Omega}$ is only $\delta g_{t\phi}$, which is given as
\begin{align}
 \delta g_{t\phi} &= r^2\left[e^{-4\varphi}\omega + e^{\nu/2}\left(e^{-4\varphi}-1\right){\cal V}\right]\sin\theta\partial_\theta P_\ell.
\end{align}
Then from the $(t,\phi)$ component of the Einstein equation (\ref{Einstein}), one can get the following equation
\begin{align}
  \bigg[&-\omega'' +\left(8\varphi'+\frac{\nu'}{2} + \frac{\zeta'}{2} - \frac{4}{r}\right)\omega' + \left\{4\varphi'' - 2\varphi' \left(8\varphi'+\nu'+\zeta'-\frac{8}{r}\right)+ \nu'' + \left(\frac{\nu'}{2} + \frac{1}{r}\right)\left(\nu'-\zeta'\right) + \frac{(\ell-1)(\ell+2)}{r^2}e^{\zeta}\right\}\omega  \nonumber \\
  &+ e^{\nu/2} \left(e^{4\varphi}-1\right)\left\{{\cal V}'' + \left(\frac{\nu'}{2} - \frac{\zeta'}{2} + \frac{4}{r}\right){\cal V}' + \left(-\frac{\nu''}{2} + \frac{\nu'\zeta'}{4} - \frac{(\nu')^2}{2} + \frac{\nu'}{r} + \frac{\zeta'}{r} - \frac{(\ell-1)(\ell+2)}{r^2}e^{\zeta}\right){\cal V}\right\}  \nonumber \\
  &+ e^{\nu/2}\left\{8\varphi'{\cal V}' + \left(4\varphi'' +2\varphi' \left(\frac{8}{r} - 8\varphi' +\nu' - \zeta'\right)\right){\cal V}\right\}\bigg]\sin\theta\partial_\theta P_\ell   \nonumber \\
  &=16\pi G e^{\zeta} \left[-\tilde{\rho}e^{-2\varphi}\omega -\left(1-e^{-4\varphi}\right) e^{\nu/2+2\varphi}\tilde{P}{\cal V} + \frac{\left(\varphi'\right)^2}{2kG}e^{-\zeta} \left\{-\omega + e^{\nu/2}\left(e^{4\varphi}-1\right){\cal V}\right\}\right] \sin\theta \partial_\theta P_\ell \nonumber \\
  &+K\bigg[-\nu'\omega' - \left\{\nu'' + \left(-4\varphi' + \frac{4}{r} - \frac{\zeta'}{2} - \frac{3\nu'}{4}\right)\nu'\right\}\omega - \nu' e^{\nu/2}{\cal V}' \nonumber \\
  &+ e^{\nu/2} \left\{-\nu'' +\left(4\varphi' - \frac{4}{r} +\frac{\zeta'}{2} +\frac{\nu'}{4}\left(1-e^{-4\varphi}\right)\right)\nu'\right\}{\cal V}\bigg]\sin\theta\partial_\theta P_\ell
    -16\pi G \left(\tilde{\rho} + \tilde{P}\right) e^{\zeta-2\varphi}\tilde{\Omega}\sin^2\theta.
\label{pEinstein-00}
\end{align}
On the other hand, the additional equation to express the relation between $\omega$ and ${\cal V}$ can be obtained from the $\phi$ component of the vector field equation (\ref{vector}), which is
\begin{align}
 K&\bigg[e^{-\nu/2}\left\{\omega''-\left(8\varphi'+\nu'+\frac{\zeta'}{2}-\frac{4}{r}\right)\omega'-\left(4\varphi''+\nu''-\left(\nu'-\frac{2}{r}+4\varphi'\right)\left(\frac{\zeta'}{2}+4\varphi'-\frac{2}{r}\right)+\frac{2}{r^2}+\frac{\ell(\ell+1)}{r^2}e^{\zeta}\right)\omega\right\}   \nonumber \\
 &+ {\cal V}'' + \left(-8\varphi'+\frac{4}{r}-\frac{\zeta'}{2}\right){\cal V}' + \left(-4\varphi'' - \frac{\nu''}{2}-\frac{\nu'}{4}\left(\nu'-\zeta'\right)+ \left(4\varphi'-\frac{2}{r}\right)\left(\frac{\zeta'}{2}+4\varphi'-\frac{2}{r}\right)-\frac{2}{r^2}-\frac{\ell(\ell+1)}{r^2}e^{\zeta}\right){\cal V}  \nonumber \\
 &+ \frac{\nu'}{2}e^{4\varphi}\left({\cal V}' + \frac{\nu'}{2}{\cal V}\right)\bigg]\frac{1}{\sin\theta}\partial_\theta P_\ell
 =8\pi G\left(1-e^{-4\varphi}\right)\left(\tilde{\rho}+\tilde{P}\right)e^{\zeta+2\varphi}\left[{\cal V}\frac{1}{\sin\theta}\partial_\theta P_\ell - e^{-\nu/2}\tilde{\Omega}\right].
 \label{pvector-00}
\end{align}
Especially, in this article we adopt the case for $\ell=1$ to reproduce the GR limit of TeVeS \cite{Hartle1967}. In fact, the other value of $\ell$ can not be chosen to satisfy the regularity condition at the stellar center and the asymptotic behavior far from the star. Then, by combining the above equations (\ref{pEinstein-00}) and (\ref{pvector-00}), one can obtain the following equations
\begin{align}
  e^{\nu/2+4\varphi}&{\cal V}'' = -\nu'\left(K-\frac{1}{2}\right)\left(\omega' +e^{\nu/2}{\cal V}'\right) - e^{\nu/2+4\varphi}\left(\nu'-\frac{\zeta'}{2}+\frac{4}{r}\right){\cal V}' + \frac{8\pi G}{K}e^{\zeta-2\varphi}\left(\tilde{\rho}+\tilde{P}\right)\left(2K-1+e^{4\varphi}\right)\tilde{\Omega} \nonumber \\
  & - \left[2\varphi'\nu'+\frac{(\nu')^2}{2} - \frac{\nu'}{r}-\frac{2\zeta'}{r}-\frac{2}{r^2}(e^{\zeta}-1) + 16\pi G e^{\zeta-2\varphi}\tilde{\rho}+\frac{8\pi}{k}(\varphi')^2 +K\left\{\nu''+\left(\frac{4}{r}-4\varphi'-\frac{\zeta'}{2}-\frac{3\nu'}{4}\right)\nu'\right\}\right]\omega \nonumber \\
  & +e^{\nu/2+4\varphi}\left[\frac{\nu''}{2}-\frac{\nu'\zeta'}{4}+\frac{(\nu')^2}{4}-\frac{\nu'}{r}-\frac{\zeta'}{r}+\frac{8\pi}{k}(\varphi')^2\right]{\cal V} 
   +e^{\nu/2}\bigg[-2\varphi'\nu' - \frac{(\nu')^2}{4}+\frac{\nu'}{r}+\frac{2\zeta'}{r}+\frac{2}{r^2}(e^{\zeta}-1) \nonumber \\
   & -16\pi Ge^{\zeta+2\varphi}\tilde{P} -\frac{8\pi}{k}(\varphi')^2 + \frac{8\pi G}{K}\left(\tilde{\rho}+\tilde{P}\right)e^{\zeta+2\varphi} + K \left\{-\nu''+\left(4\varphi'-\frac{4}{r}+\frac{\zeta'}{2}+\frac{\nu'}{4}\right)\nu'\right\}\bigg]{\cal V} \nonumber \\
   & +e^{\nu/2-4\varphi}\left[16\pi G e^{\zeta+2\varphi}\tilde{P} - \frac{8\pi G}{K}\left(\tilde{\rho}+\tilde{P}\right)e^{\zeta+2\varphi} - \frac{K}{4}(\nu')^2\right]{\cal V},
\label{pvector}
\end{align}
and
\begin{align}
  -&\omega'' +\left(8\varphi'+\frac{\nu'}{2} + \frac{\zeta'}{2} - \frac{4}{r}\right)\omega' + \left\{4\varphi'' - 2\varphi' \left(8\varphi'+\nu'+\zeta'-\frac{8}{r}\right)+ \nu'' + \left(\frac{\nu'}{2} + \frac{1}{r}\right)\left(\nu'-\zeta'\right)\right\}\omega  \nonumber \\
  &+ e^{\nu/2} \left(e^{4\varphi}-1\right)\left\{{\cal V}'' + \left(\frac{\nu'}{2} - \frac{\zeta'}{2} + \frac{4}{r}\right){\cal V}' + \left(-\frac{\nu''}{2} + \frac{\nu'\zeta'}{4} - \frac{(\nu')^2}{2} + \frac{\nu'}{r} + \frac{\zeta'}{r}\right){\cal V}\right\}  \nonumber \\
  &+ e^{\nu/2}\left\{8\varphi'{\cal V}' + \left(4\varphi'' +2\varphi' \left(\frac{8}{r} - 8\varphi' +\nu' - \zeta'\right)\right){\cal V}\right\}  \nonumber \\
  &=16\pi G \left(\tilde{\rho} + \tilde{P}\right) e^{\zeta-2\varphi}\tilde{\Omega} +16\pi G e^{\zeta} \left[-\tilde{\rho}e^{-2\varphi}\omega -\left(1-e^{-4\varphi}\right) e^{\nu/2+2\varphi}\tilde{P}{\cal V} + \frac{\left(\varphi'\right)^2}{2kG}e^{-\zeta} \left\{-\omega + e^{\nu/2}\left(e^{4\varphi}-1\right){\cal V}\right\}\right] \nonumber \\
  &-K\bigg[\nu'\omega' + \left\{\nu'' + \left(-4\varphi' + \frac{4}{r} - \frac{\zeta'}{2} - \frac{3\nu'}{4}\right)\nu'\right\}\omega + \nu' e^{\nu/2}{\cal V}'   - e^{\nu/2} \left\{-\nu'' +\left(4\varphi' - \frac{4}{r} +\frac{\zeta'}{2} +\frac{\nu'}{4}\left(1-e^{-4\varphi}\right)\right)\nu'\right\}{\cal V}\bigg].
\label{pEinstein}
\end{align}
Note that the GR limit of Eq. (\ref{pEinstein}) agrees with the well-known equation describing the frame dragging in GR. With these two equations (\ref{pvector}) and (\ref{pEinstein}) and with the appropriate boundary conditions, the distributions of $\omega(r)$ and ${\cal V}(r)$ can be determined. 
Additionally, with an asymptotic flatness, one can show that $\omega$ and ${\cal V}$ are decreasing as $1/r^3$ far from the central object.

\subsection{Numerical Results}
\label{sec:III-2}

\begin{figure}[htbp]
\begin{center}
\includegraphics[scale=0.45]{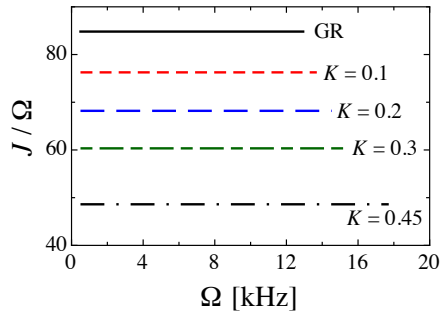} 
\end{center}
\caption{
Dependence of $J/\tilde{\Omega}$ on the angular velocity $\tilde{\Omega}$ for the stellar model with EOS A and $M_{\rm ADM}=1.4M_\odot$, where the solid line corresponds to the case of GR while the broken lines are results in TeVeS with different values of $K$.
}
\label{fig:Omega-J}
\end{figure}

\begin{figure}[htbp]
\begin{center}
\includegraphics[scale=0.45]{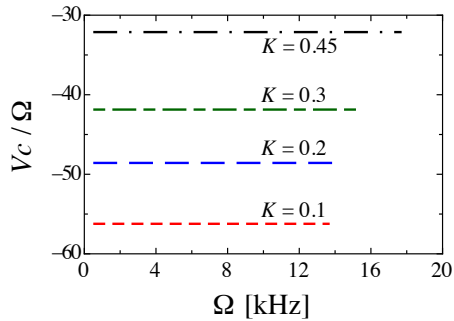} 
\end{center}
\caption{
Dependence of ${\cal V}_c/\tilde{\Omega}$ on the angular velocity $\tilde{\Omega}$ for the stellar model with EOS A and $M_{\rm ADM}=1.4M_\odot$.
}
\label{fig:Omega-Vc}
\end{figure}

In order to determine the distributions of $\omega(r)$ and ${\cal V}(r)$ with the fixed value of $K$ for the adopted stellar model, we impose two boundary conditions, i.e., the regularity condition at the stellar center and the asymptotic flatness far from the star. In practice, at stellar center we set that $\omega(0)=\omega_0$, $\omega'(0)=0$, ${\cal V}(0)={\cal V}_0$, and ${\cal V}'(0)=0$, where $\omega_0$ and ${\cal V}_0$ are some constants. Then we find the correct values of $\omega_0$ and ${\cal V}_0$ in such a way that the  solutions of $\omega(r)$ and ${\cal V}(r)$ should satisfy the asymptotic behavior as mentioned in the previous section, i.e., with the trial values of $\omega_0$ and ${\cal V}_0$ the above ordinary differential equations can be integrated outward and we search the correct values by changing the trial values iteratively, where for integration we use the 4th order Runge-Kutta method. In this way, we can determine numerically the distributions of $\omega(r)$ and ${\cal V}(r)$ and those behaviors far from the central star can be described as
\begin{align}
  \omega(r) &= \frac{2J}{r^3} + {\cal O}\left(\frac{1}{r^4}\right), \\
  {\cal V}(r)  &= \frac{2{\cal V}_c}{r^3} +  {\cal O}\left(\frac{1}{r^4}\right).
\end{align} 
It should be noticed that in GR the distribution of $\omega(r)$ outside the star can be described analytically as $\omega(r) = 2J/r^3$, where the constant $J$ is corresponding to the total angular momentum of the star \cite{Hartle1967}. Additionally, it is well known that the angular momentum in GR is linearly related to the angular velocity for slow rotation as $J=I\tilde{\Omega}$, where the constant of proportionality $I$ corresponds to the relativistic generalization of momentum of inertia for slowly rotating systems \cite{Hartle1967}. While, in the case of TeVeS, we can see the same feature as the case of GR, i.e., as shown in Figs. \ref{fig:Omega-J} and \ref{fig:Omega-Vc}, the values of $J/\tilde{\Omega}$ and ${\cal V}_c/\tilde{\Omega}$ are independent of the value of angular velocity $\tilde{\Omega}$ if only one chooses the non-rotating stellar models.

%

\begin{figure}[htbp]
\begin{center}
\begin{tabular}{cc}
\includegraphics[scale=0.45]{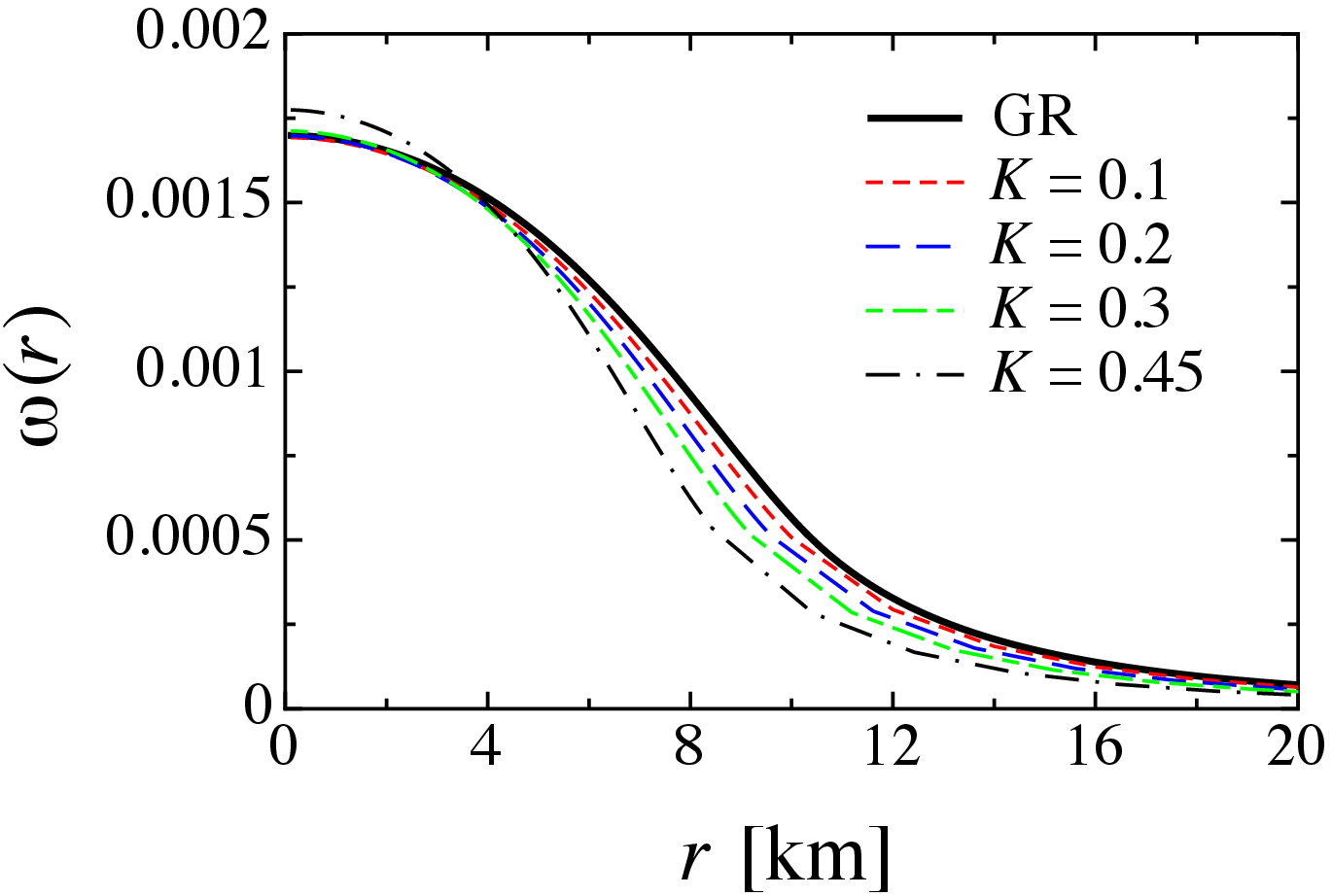} &
\includegraphics[scale=0.45]{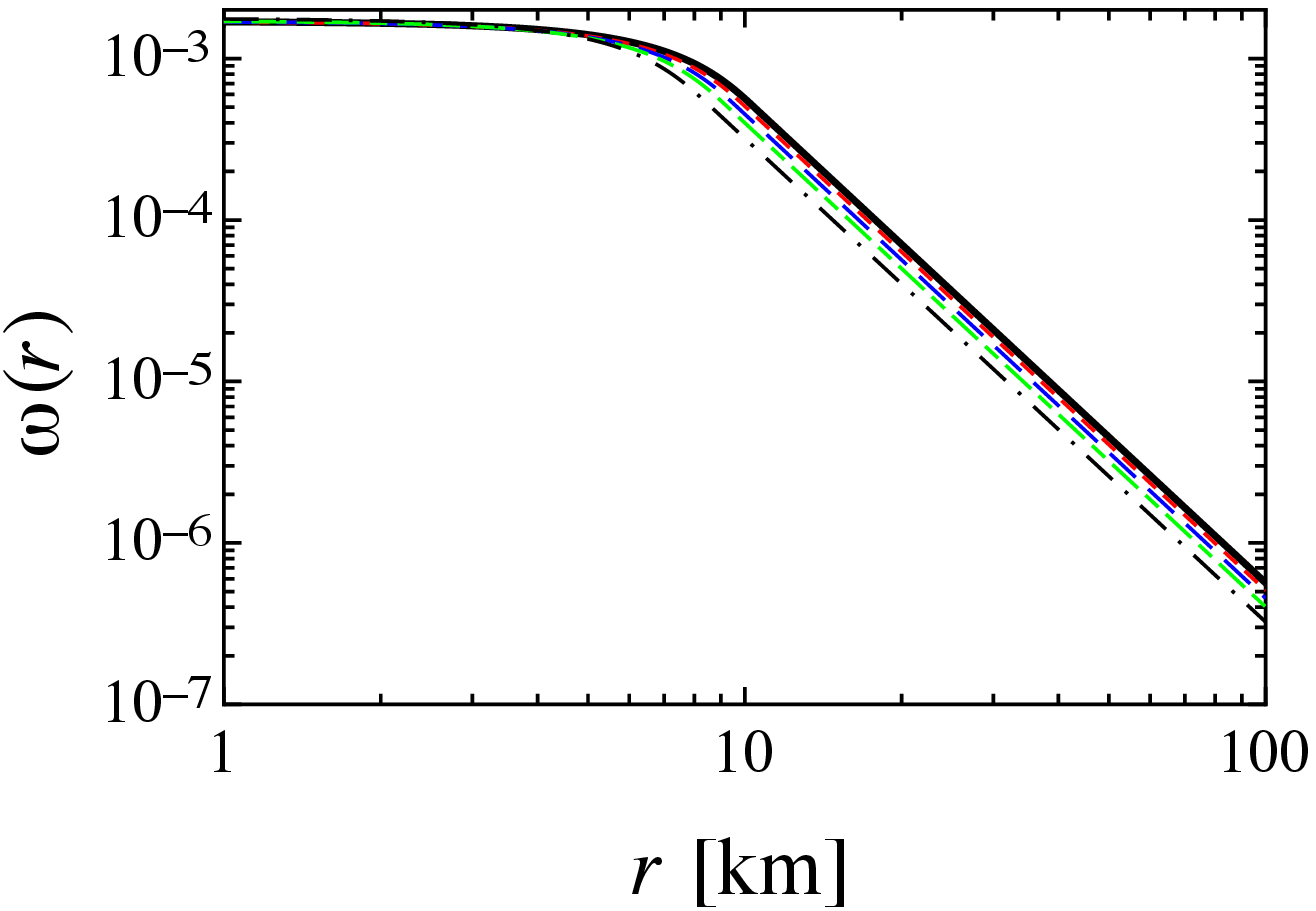} \\
\end{tabular}
\end{center}
\caption{
Distributions of $\omega(r)$ with different values of $K$ for the stellar model with EOS A, $M_{\rm ADM}=1.4M_\odot$, and $\tilde{\Omega}=1$ kHz. In order to compare the results, the solid line shows the distribution in GR.
}
\label{fig:r-omega}
\end{figure}

\begin{figure}[htbp]
\begin{center}
\begin{tabular}{cc}
\includegraphics[scale=0.45]{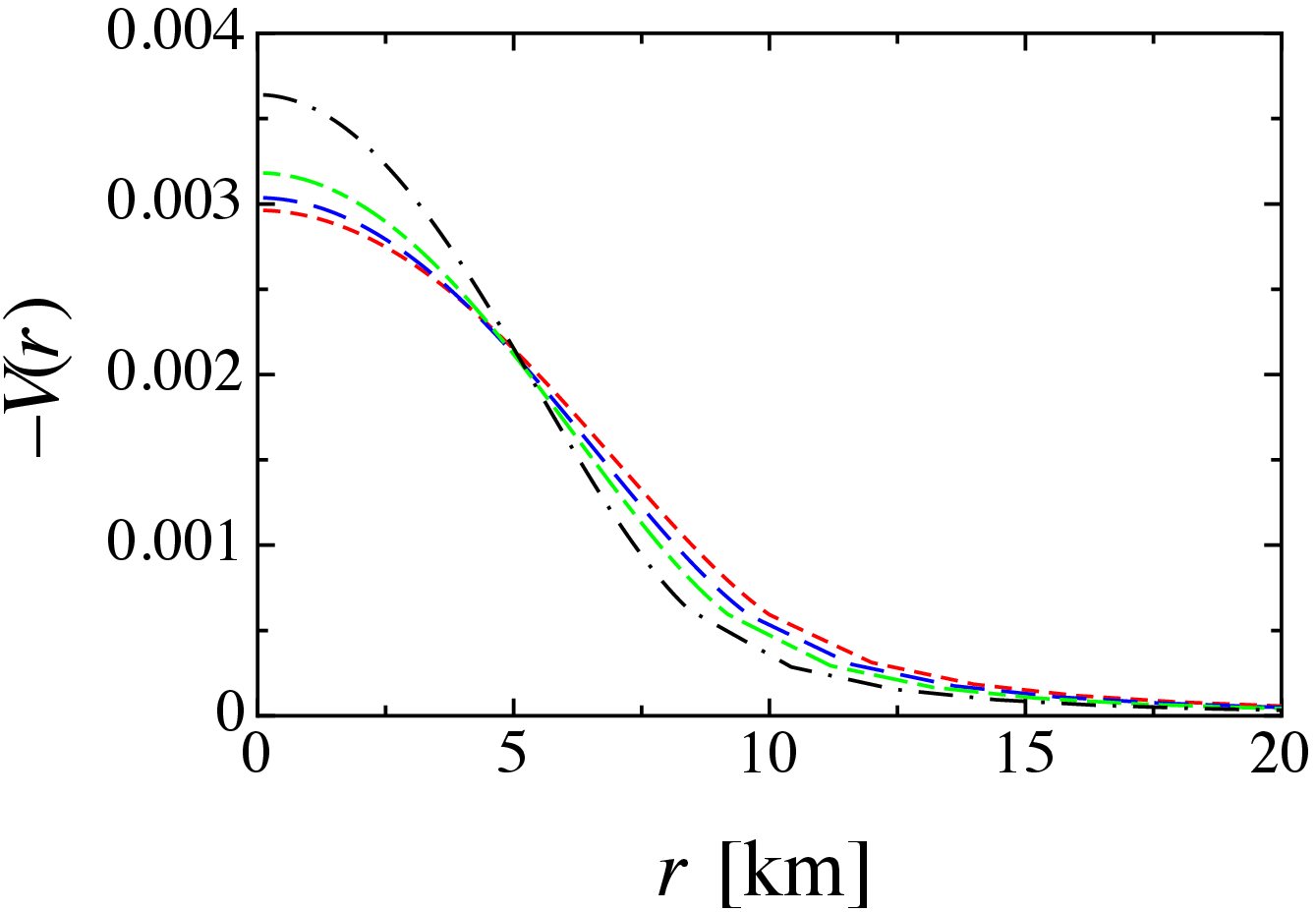} &
\includegraphics[scale=0.45]{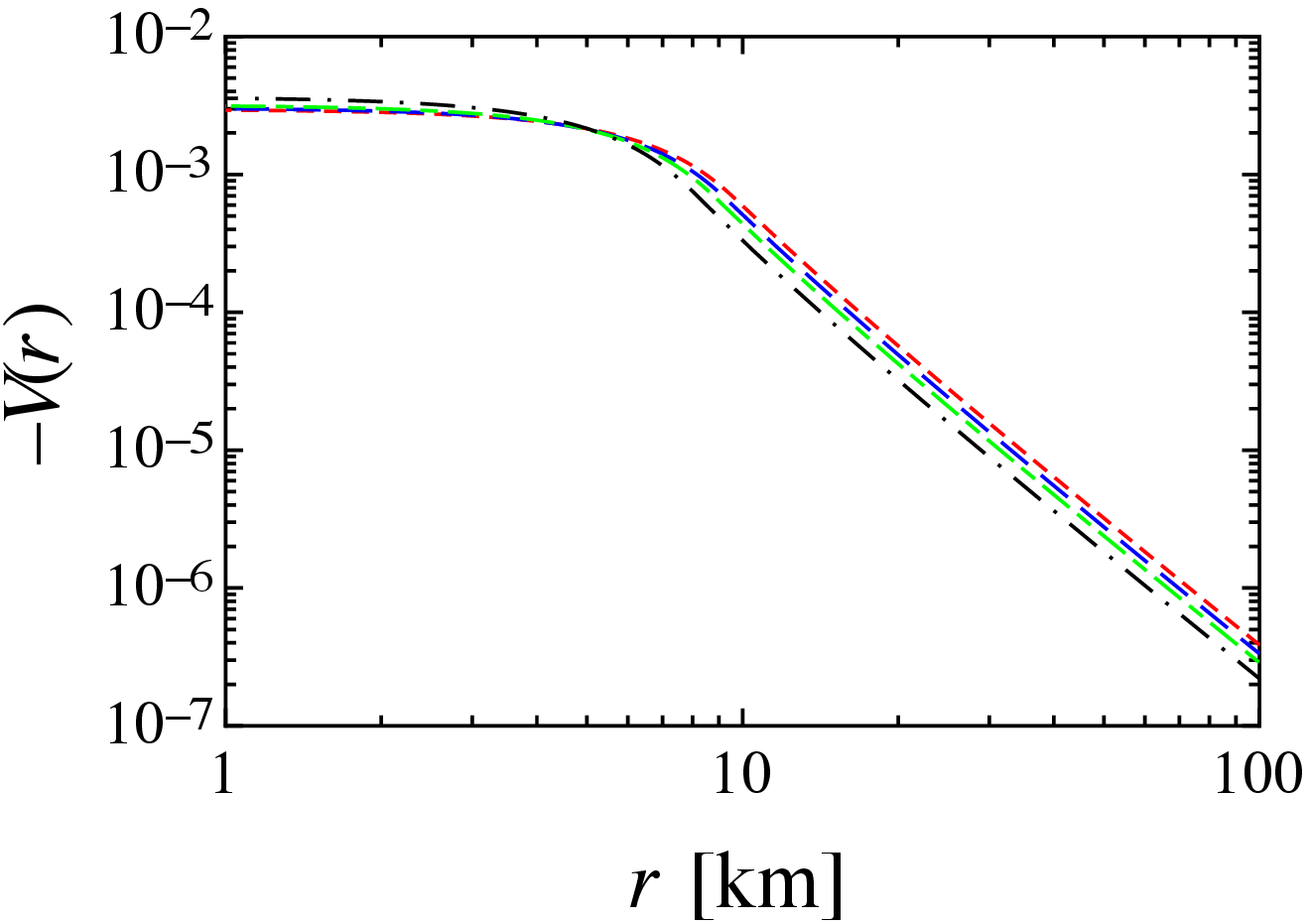} \\
\end{tabular}
\end{center}
\caption{
Distributions of ${\cal V}(r)$ with different values of $K$ for the stellar model with EOS A, $M_{\rm ADM}=1.4M_\odot$, and $\tilde{\Omega}=1$ kHz, where the different lines correspond to the results with values of $K$ shown in Fig. \ref{fig:r-omega}.
}
\label{fig:r-V}
\end{figure}

Fig. \ref{fig:r-omega} shows the distributions of $\omega(r)$ with different values of $K$ for the stellar model with EOS A, $M_{\rm ADM}=1.4M_\odot$, and $\tilde{\Omega}=1$ kHz. At a glance, we can recognize the difference of distributions between in GR and in TeVeS. Especially, with higher value of $K$ those differences become obvious even in the interior of the star. As mentioned before, the behavior of $\omega(r)$ far from star is proportional to $r^{-3}$, which can be seen in the right panel of Fig. \ref{fig:r-omega}. On the other hand, Fig. \ref{fig:r-V} shows the distributions of ${\cal V}(r)$ with different values of $K$ for the same stellar model as in Fig. \ref{fig:r-omega}. In the right panel of Fig. \ref{fig:r-V} one can see the behavior of ${\cal V}(r)$ far from star.
From Figs. \ref{fig:r-omega} and \ref{fig:r-V}, we can find the important point that as the value of $K$ becomes smaller, the distribution of $\omega(r)$ in TeVeS approaches to that in GR while the induced vector field ${\cal V}(r)$ does not vanish. Namely, with smaller value of $K$ the physical metric in TeVeS is almost same as that in GR, but there still exists the additional vector fields induced by the stellar rotation. So if one will consider the stellar oscillations and/or the emitted gravitational waves in TeVeS, this induced vector field could play a role as source term in the linearized equations which is corresponding to the linearized right hand side in Eq. (\ref{Einstein}), and as a result the deviation in frequencies depending on the gravitational theory can be seen even if the physical metric is not so different from each other. Thus such kinds of observation could tell us the gravitational theory in the strong field regime even if the value of $K$ is quite small.

%

\begin{figure}[htbp]
\begin{center}
\includegraphics[scale=0.45]{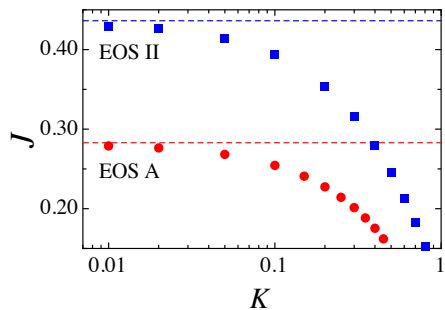} 
\end{center}
\caption{
Dependence of $J$ on $K$ for the stellar models with $M_{\rm ADM}=1.4M_\odot$ and $\tilde{\Omega}=1$ kHz for EOS A and EOS II. The horizontal broken lines are corresponding to the values for each EOS in GR.
}
\label{fig:K-J-1000Hz}
\end{figure}

\begin{figure}[htbp]
\begin{center}
\includegraphics[scale=0.45]{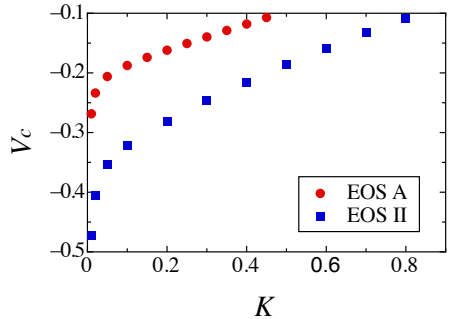} 
\end{center}
\caption{
Dependence of ${\cal V}_c$ on $K$ for the stellar models with $M_{\rm ADM}=1.4M_\odot$ and $\tilde{\Omega}=1$ kHz for EOS A and EOS II. 
}
\label{fig:K-Vc-1000Hz}
\end{figure}

Furthermore, the dependences of $J$ and ${\cal V}_c$ on the value of $K$ are shown in Figs. \ref{fig:K-J-1000Hz} and \ref{fig:K-Vc-1000Hz}, where the stellar masses are fixed to be $M_{\rm ADM}=1.4M_\odot$ and the angular velocity are adopted that $\tilde{\Omega}=1$ kHz for EOS A and EOS II. In Fig. \ref{fig:K-J-1000Hz} the horizontal broken lines denote the values of $J$ in GR for each EOS. From this figure, we can see that depending on the value of $K$, the angular momentum $J$ could become 42.7\% smaller for EOS A and 65.0\% smaller for EOS II than those expected in GR. Even if this difference cannot be directly observed, one could find a kind of evidence depending on the gravitational theory in the strong field regime by observing the phenomena around rotating compact object.

%

%

\section{Conclusion}
\label{sec:IV}

In this article, in order to examine the rotational effect around the neutron star in the tensor-vector-scalar (TeVeS) theory, we consider the slowly rotating relativistic stars with a uniform angular velocity $\tilde{\Omega}$. To deal with this problem in TeVeS, one has to take into account not only the usual frame dragging $\omega$ but also the induced vector field ${\cal V}$, which is corresponding to the $\phi$ component.
The equations for $\omega(r)$ and ${\cal V}(r)$ are derived from the Einstein and vector field equations, and then the distributions of those variables are determined numerically with appropriate boundary conditions.

As a result, we find that, similar to the case in GR, the value of $J/\tilde{\Omega}$ in TeVeS is constant if the stellar mass and value of $K$ are fixed, where $J$ corresponds to the angular momentum. Additionally, with higher value of $K$, the distribution of $\omega(r)$ deviates obviously from that in GR due to the existence of induced vector field and this deviation can be seen even in the interior region of star. On the other hand with smaller value of $K$, although $\omega$ approaches to the that in GR, the induced vector field dose not vanish. That is, even if the stellar properties in TeVeS with small value of $K$, such as the mass and radius, are almost same as those in GR, the induced vector field in TeVeS could still exist. This is a crucial difference depending on the gravitational theory. Thus through the observable phenomena, such as  the stellar oscillation and emitted gravitational waves, it is possible to distinguish the gravitational theory in the strong field regime. Furthermore, through the stellar magnetic effect, one might be see another effect coupling to the induced vector field. For example, some quasi-periodic oscillations are observed recently in the giant flares and these are believed to be related to the oscillations of strong magnetized neutron stars \cite{Sotani2007}. Considering the stellar magnetic fields, one might be possible to obtain the further constraint in the theory.

\acknowledgments
We thank Demetrios Papadopoulos, Kostas D. Kokkotas, Miltos Vavoulidis, and Erich Gaertig for valuable comments, and also the referees for fruitful comments.  This work was supported via the Transregio 7 ``Gravitational Wave Astronomy" financed by the Deutsche Forschungsgemeinschaft DFG (German Research Foundation).



\end{document}